**Magnetism-Inspired Quantum-Mechanical Model of Gender Fluidity**

Ivan S. Maksymov

*Artificial Intelligence and Cyber Futures Institute, Charles*
*Sturt University, Bathurst, NSW 2795, Australia*

imaksymov@csu.edu.au

Abstract

Quantum-mechanical models of human cognition, opinion formation and decision-making have changed the way we understand and predict human behaviour in many practical situations, including political elections, financial decisions and international affairs. Yet, at present, such models overlook certain essential social aspects of human behaviour and self-identification. In this paper, we introduce a magnetism-inspired quantum-mechanical model of gender fluidity, a concept that challenges social norms across the globe. Addressing a number of independent suggestions made by members of the general public concerning a potential analogy between quantum superposition and non-binary self-identification, we explore new territories, demonstrating that physic of magnetism can help explain gender fluidity and similar social phenomena better than the traditional quantum-mechanical models of human cognition and perception. We anticipate that the proposed model can be used to analyse experimental datasets aimed to develop sexual orientation and gender identity legal definitions as well as to create artificial intelligence systems that can sensibly identify both binary and non-binary genders.

## I.  INTRODUCTION

To survive and evolve in this world, humans have always strove to understand who and what they are. This continuous process has led to the creation of certain norms, acceptable social roles and behaviour patterns. One of such norms is the concept of gender, a set of socially constructed characteristics of women, men, girls and boys that varies from society to society and can change over time [1–7], playing an important role in popular culture and literature [8, 9].

The concept of gender fluidity challenges social norms [10–17]. Adopting a binary point of view, being gender fluid can be defined as having a different gender identity at different times [18]. For example, at one moment an individual may identify themselves as female and at another moment they may identify themselves as male. Yet, an individual may also identify themselves as both male and female at the same time or none. Such identity shifts can happen at different timescale: several times a day, weekly, monthly or yearly [19]. In fact, gender fluidity is more complex. However, the extent of its complexity has not been established yet despite the attempts to understand it using the methods adopted in the domain of complex systems [20].

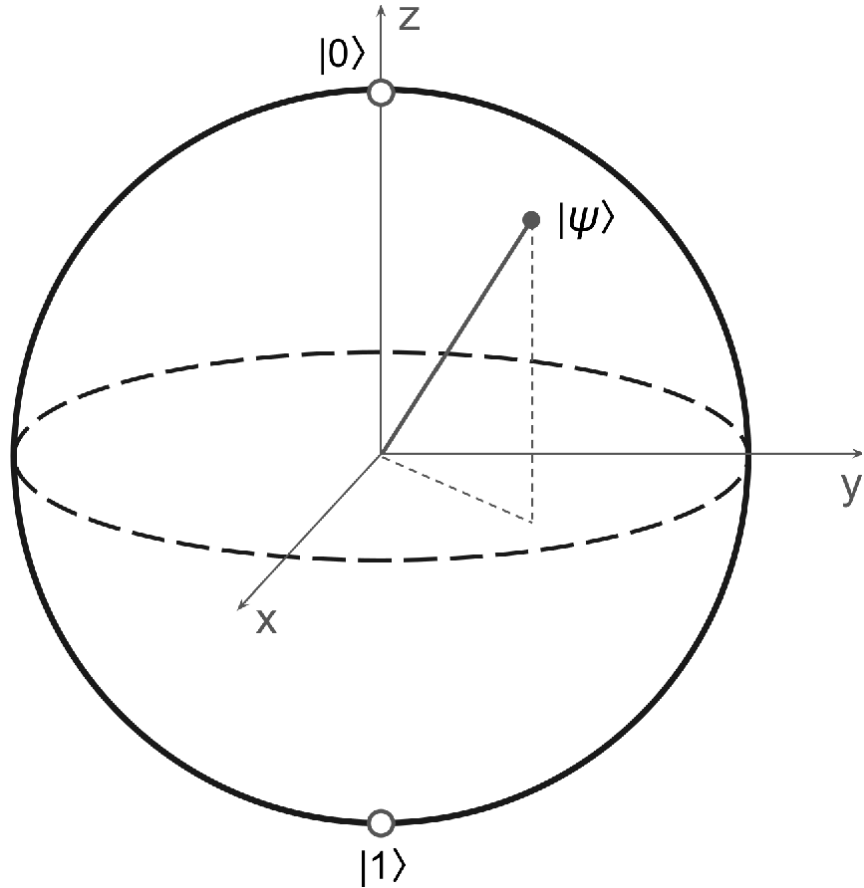

FIG. 1. Illustration of a projective measurement of a qubit $|\psi\rangle$ using the Bloch sphere.

Complex systems is an umbrella term applied to a methodological approach used in physics, engineering, life and social sciences, management and health to reveal how relationships between parts give rise to a collective behaviours of the entire system, also explaining how the system interacts with the environment [21, 22]. The human brain is also a complex system [23] (this explains why neuroscience can also help understand the origins of gender fluidity [13]). Moreover, the brain is a nonlinear dynamical system since its behaviour changes over time [24]. Thus, due to complexity of gender fluidity and a dynamical nature of the brain and social processes relevant to gender fluidity, physical and mathematical principles underpinning the nonlinear dynamics [25] could be used to create a viable model of gender fluidity.

Indeed, using the fundamental principles of physics we can comprehend and challenge the common binary view of male and female through the prism of our basic understanding of nature. For example, let us consider a traditional digital computer and a quantum computer [26]. Similarly to an on/off light switch, a bit of a digital computer is always in one of two physical states corresponding to the binary values '0' and '1'. However, a quantum computer uses a quantum bit (qubit) that can be in states $|0\rangle = [1\ 0]$ and $|1\rangle = [0\ 1]$. These states are analogous to the '0' and '1' binary states of the classical digital computer. However, a qubit also exists in a continuum of states between $|0\rangle$ and $|1\rangle$, i.e. its states are a superposition $|\psi\rangle = \alpha|0\rangle + \beta|1\rangle$ with $|\alpha|^2 + |\beta|^2 = 1$.

Computational algorithms based on measurements of the states of a qubit are exponentially faster than any possible deterministic classical algorithm [26]. Subsequently, it has been demonstrated that quantum mechanics can model human mental states better than any existing classical model [27–30]. In particular, it was suggested that a quantum-like superposition of human mental states can explain human preference, anomalies of decision-making and perception of optical illusions [31–41]. It is noteworthy that the models proposed in the cited papers are mostly phenomenological, i.e. they describe the psychological and behavioural science phenomena without necessarily conforming to the existing theories. Despite perceived limitations that are yet to be investigated in more detail, this approach has proven useful for analysis of complex experimental data, serving as a valuable tool for researchers working across several disciplines [42–44].

Importantly, quantum-mechanical models go hand in hand with the general public interest in gender fluidity. Indeed, it was suggested that, since the quantum physics that describes the universe is not binary, the gender must also be non-binary [45–47]. Often referring to the famous book [48], several authors elaborated this idea and tried to establish a stronger link between science and gender (see, e.g., [49, 50]).

From the physical point of view, the ideas expressed in the works cited above may be illustrated using the concept of the Bloch sphere (Fig. 1). When a quantum measurement is done [26, 51], a closed qubit system interacts in a controlled way with an external system, thereby revealing the state of the qubit under measurement. Using the projective measurement operators $M_0 = |0\rangle\langle 0|$ and $M_1 = |1\rangle\langle 1|$ [26], the measurement probabilities for $|\psi\rangle = \alpha|0\rangle + \beta|1\rangle$ are $P_{|0\rangle} = |\alpha|^2$ and $P_{|1\rangle} = |\beta|^2$, which means that the qubit will be in one of its basis states. Visually, the measurement procedure means that the qubit is projected on one of the coordinate axes (e.g., $z$-axis in Fig. 1). Thus, we may assume that the state $|\psi\rangle$ corresponds to the non-binary gender and that the quantum measurement makes a probabilistic prediction of whether the non-binary gender state is closer to the purely binary male or female gender.

However, while this theoretical approach serves as a formal model of the intuitive suggestions made in [45–47] and the relevant works, measuring a quantum system results in a collapse of the superposition quantum state that describes that system into one definite state, which is an essential feature of quantum mechanics [51]. Of course, on the level of the model, this peculiarity of quantum theory does not mean that the actual gender state is destroyed by the measurement. Nevertheless, despite the fact that quantum physics has achieved experimental success and wide-range applicability, the debate about the interpretation of the quantum measurement continues on a more philosophical level [52], complicating the comprehension of social quantum-mechanical models by non-experts in physics.

In this paper, we propose a more intuitive quantum-physical model of gender fluidity that helps avoid using the concept of quantum measurement. Figure 2a, b schematically illustrates how gender fluidity can be described using a combination of the concepts of binarity and magnetisation. While the depiction of the binary and non-binary genders in Fig. 2a is used rather for illustration purposes (the readers interested in a more rigorous picture are referred to [19, 53, 54]), the arrows in Fig. 2b, despite their schematic character, have a clear physical meaning—they correspond to the macrospins. The direction of the leftmost and rightmost arrows are assigned to the binary definitions of the gender. The directions of the in-between arrows phenomenologically describe non-binary gender identities.

Spin is a quantum-mechanical concept that is pivotal for magnetism [51, 55]. Spin cannot be explained using the principles of classical physics. However, a theoretical relationship between spin and classical rotation can be established [56] (see Fig. 3a, b). Yet, one can create a feasible classical physical model of magnetism using the concept of macrospin [57]. When in a substance the number of spins pointing in different directions is equal, the magnetic properties of this substance are cancelled. However, when most of the spins point in the same direction forming a macrospin, the substance becomes magnetic. This substance can be further magnetised using a static magnetic field to form a magnet.

The direction of the macrospin can be forced to change when the magnetised substance is affected by a dynamical (time-varying) magnetic field or an electric current. When the strength of the forcing is low, the macrospin just slightly deviates from its original orientation and then quickly returns to the equilibrium position (Fig. 3c.i). As the strength of the forcing increases, the direction of macrospin significantly deviates from the equilibrium position and the macrospin starts to precess around the direction of the applied static magnetic field (Fig. 3c.ii). We can visualise this precession as the motion of a spinning top, as illustrated in Fig. 3b. Importantly, the macrospin can continue precessing for an indefinitely long time due to a balance between the forcing and damping processes in the substance. Furthermore, when a strong dynamic magnetic field or electric current is applied, the amplitude of the precession becomes high enough for the spin to permanently change its original direction for the opposite one via the process of magnetisation reversal (Fig. 3c.iii).

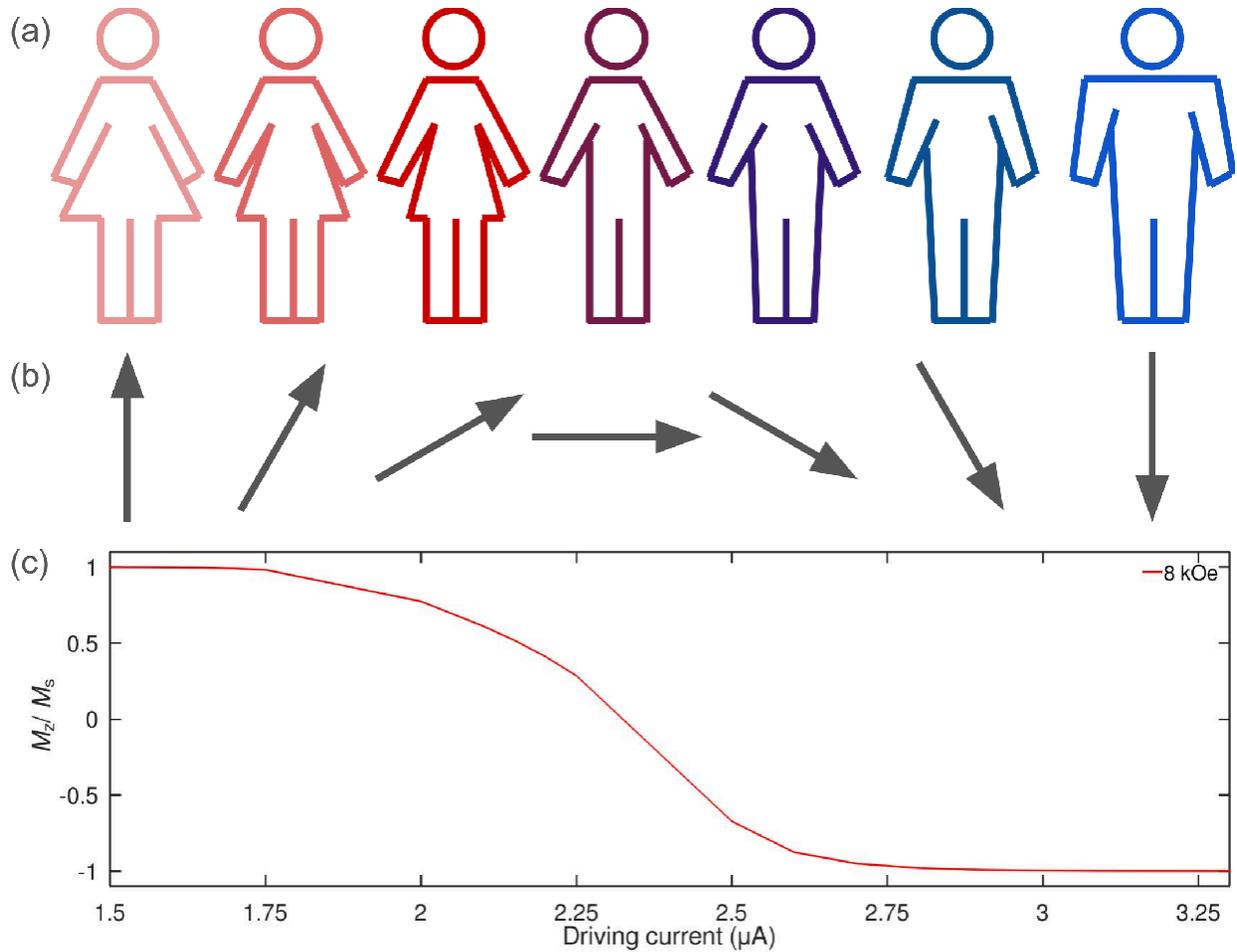

FIG. 2. (a, b) Illustration of how gender fluidity can be described using a combination of the concepts of binarity and the physical process of magnetisation. The arrows correspond to the macrospins. The direction of the leftmost and rightmost arrows denote the binary gender. The directions of the in-between arrows describe non-binary gender identities. (c) Result of a numerical simulation of magnetisation reversal for the static magnetic field $H_0 = 8$ kOe. The left- and right- most macrospin arrows in Panel (b) correspond to the values 1 and $-1$, respectively.

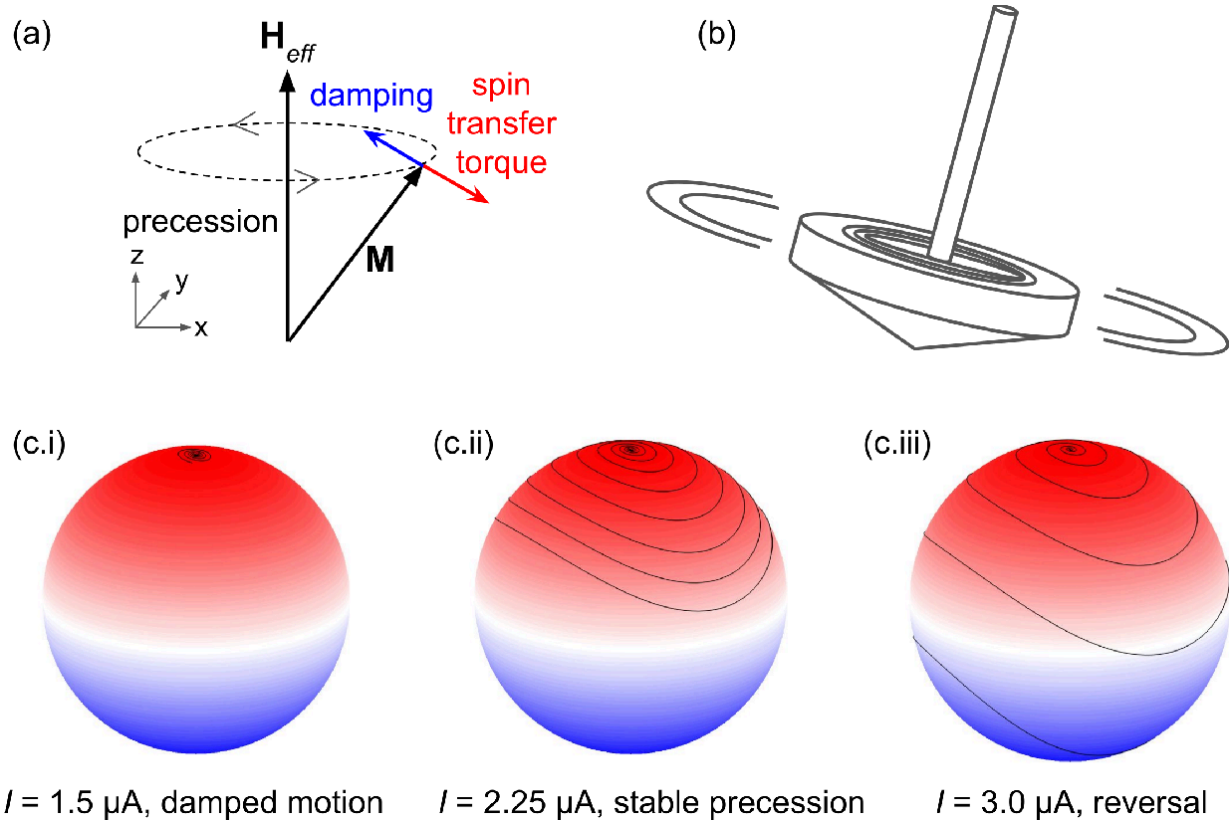

(a)

**H**_{eff}

damping

spin
transfer
torque

precession

z
y
x

**M**

(b)

(c.i)

(c.ii)

(c.iii)

$I$ = 1.5 μA, damped motion    $I$ = 2.25 μA, stable precession    $I$ = 3.0 μA, reversal

FIG. 3. (a) Illustration of precession of the magnetisation vector **M** representing a macrospin. (b) The movement associated with the precession of **M** can be compared with the wobbling motion of a spinning top. Simulated results: (c.i) **M** spirals back toward the direction of **H**_{eff} due to the damping, (c.ii) stable precession of **M** around **H**_{eff} and (c.iii) reversal of the direction of **M**. In the model, the applied field is $H_0$ = 8 kOe. The coloured sphere is used for visualisation only.

## II. METHODS: MAGNETISATION REVERSAL MODEL

To demonstrate the reversal of the macrospin direction and represent the non-binary gender states illustrated in Fig. 2, we create a rigorous physical model of magnetisation dynamics in a ferromagnetic metal (FM) nanostructure [58]. The readers not interested in these specific physical aspects of the model can skip reading this section without affecting their understanding of the mainstream discussion.

It is known that an electric current is unpolarised since it involves electrons with a random polarisation of spins. However, when a current passes through a thin FM film with a fixed magnetisation direction it can become spin-polarised since spins become oriented predominantly in the same direction [59].

This physical effect is exploited in a spin transfer torque nano-oscillator (STNO) device that consists of a layer with a fixed direction of magnetisation **M** separated from a thinner FM layer by a non-magnetic metal layer [60]. When a spin-polarised current flows from the "fixed" magnetisation layer to the "free" magnetisation layer, the equilibrium orientation of magnetisation in the "free" layer becomes destabilised. Depending on the strength of the electric current, the destabilisation can lead to either stable precession of magnetisation of the "free" layer about the direction of the effective magnetic field or to a complete reversal of the magnetisation direction (see Fig. 3c and the discussion around it).

We solve the Landau-Lifshitz-Gilbert equation to model the dynamics of magnetisation [61]:

$$\partial \mathbf{M}/\partial t = \gamma \left[ \mathbf{H}_{eff} \times \mathbf{M} \right] + \mathbf{T}_G + \mathbf{T}_{SB} , \tag{1}$$

where $\gamma$ is the gyromagnetic ratio. The first term of the right-hand side of Eq. (1) governs the precession of $\mathbf{M}$ (Fig. 3a) about the direction of the effective magnetic field $\mathbf{H}_{eff} = H_0 \mathbf{e}_z + \mathbf{H} + \mathbf{H}_{ex}$, where $H_0$ is the external static magnetic field orientated along the $z$-axis, $\mathbf{H}$ is the dynamic field due to currents and magnetic sources such as demagnetising field and eddy current fields and $\mathbf{H}_{ex}$ is the exchange field [62]. The dissipative torque is [61]

$$\mathbf{T}_G = \alpha_G M_S^{-1} \left[ \mathbf{M} \times \partial \mathbf{M}/\partial t \right] , \tag{2}$$

where $M_s$ is the saturation magnetisation of the "free" magnetisation layer and $\alpha_G$ is Gilbert damping parameter [61, 62]. The Slonczewski-Berger torque is

$$\mathbf{T}_{SB} = \sigma_0 I M_S^{-1} \left[ \mathbf{M} \times \left[ \mathbf{M} \times \hat{e}_p \right] \right] , \tag{3}$$

where $I$ is the strength and $\hat{e}_p$ is the direction of the spin polarisation of the current. The parameter $\sigma_0$ accounts for the fundamental physical constants and the thickness of the "free" magnetisation layer [61]. Equation (1) is numerically solved using a finite-difference time-domain method and a typical set of material parameters used to model in STNO devices [62].

## III. RESULTS

We first demonstrate that the rigorous model of magnetisation reversal validates our earlier discussion of the microspin direction variation (Fig. 2b). In Fig. 2c we plot the dependence of the direction of $\mathbf{M}$ in the "free" layer as a function of the electric current strength for the applied magnetic field $H_0 = 8$ kOe (from the physical point of view, we plot the dynamics of the $M_z$ component of the magnetisation vector normalised to the saturation magnetisation $M_s$; $H_0 = 8$

kOe is a typical field strength achievable with a laboratory electromagnet).

The trajectories of $\mathbf{M}$ for the static applied magnetic field $H_0 = 8$ kOe computed at the simulated time of 1 ns are shown in Fig. 3c. In Panel (c.i), since the current strength $I = 1.5$ $\mu$A is low, the macrospin $\mathbf{M}$ spirals back toward the direction of the effective magnetic field $\mathbf{H}_{eff}$ due to the effect of damping. In Panel (c.ii), the application of a stronger current $I = 2.25$ $\mu$A results in a stable precession of $\mathbf{M}$ around $\mathbf{H}_{eff}$ achieved when the spin transfer torque compensates the damping. Finally, in Panel (c.iii) the complete reversal of the direction of $\mathbf{M}$ is observed at $I = 3$ $\mu$A.

Thus, we can see that the leftmost and right macrospin arrows in Fig. 2b correspond to the values 1 and −1 in Fig. 2c, respectively. We can assign these states to the basis binary states, defined in terms of both $|0\rangle$ and $|1\rangle$ states in Fig. 1 and binary gender roles illustrated in Fig. 2a. Furthermore, we assign the values of magnetisation in between 1 and −1 to non-binary gender identifications, noting that, unlike the arrow-based representation in Fig. 2a, the variation of the magnetisation from 1 and −1 is a gradual and continuous process.

It is noteworthy that by changing the value of $H_0$ in a technically feasible range we can control the shape of the magnetisation reversal curve. This property can be used to account for the fact that the gender fluidity picture shown in Fig. 2 is an idealisation but the real-life picture is more complex. Moreover, while it has not been established yet which curve would better fit a real-life gender fluidity scenario, we can reasonably assume that that curve would have step-like features. The proposed model can capture step-like features provided that the values of both $H_0$ and electric current $I$ are continuously varied in the modelling process. However, for the sake of clarity, in the following we will focus only on a gradually varying shape of the magnetisation reversal curve, demonstrating that a gradual variation has important implications.

## IV.  DISCUSSION

The gradual variation of the magnitude and direction of the magnetisation results in the formation of a sigmoid-like curve shown in Fig. 2c. Although the illustration of gender fluidity in Fig. 2a and the macrospin picture Fig. 2b were intuitively designed following a sigmoid-like trajectory, a sigmoidal character of the result presented in Fig. 2c has a solid scientific meaning. Indeed, it is well-established that the sigmoid function and its modifications play a central role virtually in all areas of fundamental and applied research [63–65], including the studies aimed to reveal how languages [66] and culture [67] change with time as well as how new ideas are born and spread in the society [68]. Yet, this function also fits many theoretical intuitions [69] and experimental datasets [69–73] obtained in the fields of psychology, economics, human behaviour and decision-making.

Another potential link between the physical model proposed in this paper and gender fluidity can be established based on a recent experiment that demonstrated that a perceptual illusion of having an opposite-sex body can be associated with gender fluidity [14]. Intriguingly, optical illusions have also been associated with quantum-mechanical effects [28, 31, 36, 40, 74] and they can be both modelled using the concept of qubit (Fig. 1) and the process of magnetisation reversal introduced above (for details we refer the interested reader to [74] and references therein).

Yet, gender fluidity might be understood in the framework of the concept of quantum Darwinism, a theory that explains the emergence of the classical world from the quantum world following a process of Darwinian natural selection [75, 76]. The quantum Darwinism theory clarifies the nature of quantum-classical correspondence and its postulates have been confirmed in a recent experiment [77]. These findings correlate with the works that have established a link between Darwinian natural selection, evolution and gender diversity [78–80].

On a general note, any model is an approximation that is valid mostly in one specific situation and can be applied, with certain limitations, to a number of similar situations. Yet, the sole purpose of a model can be to create a precedent for using certain theoretical approaches in a new area, motivating further research and development that, in turn, would produce a more practicable model. To a large extent the model proposed in this paper intends to create such a precedent, also urging all experts in the interdisciplinary field of gender studies to embrace novel approaches to the conduct of their research work.

## V. CONCLUSIONS

To summarise, we proposed a magnetism-inspired quantum-mechanical model of gender fluidity that has the following characteristics that differentiate it from the existing classical and quantum models of human cognition, perception and decision-making:

- Compared with the quantum models of human cognition that exploit the physical and mathematical properties of a qubit, the magnetism-inspired model is more intuitive and easier to use by experts who want to abstract from complex physical aspects. The accessibility of the model to non-experts is essential as evidenced by the previous examples, including a successful prediction of the result of major elections by sociologists who used a physical model as a black box [81, 82];

- A sigmoid-like output of the model fundamentally fits many processes that govern the dynamics of natural phenomena, societal changes and human cognition. Significant experimental evidence speaking in favour of sigmoidal approximation of human beliefs and actions has been recently produced in [83];

- The model is linked to the quantum models of perception of optical illusions, which, intriguingly, connect it to a hypothesis claiming that gender fluidity may be explained by illusory body-sex changes.

Thus, the proposed model can be used to analyse experimental datasets such as those used to develop a sexual orientation and gender identity legal index [84]. It can also lay the foundation for advanced computer algorithms that would judiciously combine physics, statistics and machine learning techniques [74]. In particular, such algorithms can be used to create a deep learning system that can correctly recognise individuals of binary and non-binary gender [85] within the framework of the law system in a democratic state [86]. Finally, the general aspects of the model can be incorporated into school instruction materials aimed to teach the student about gender diversity.

**Acknowledgements**: The author would like to thank Professor Ganna Pogrebna for valuable discussions of the magnetisation reversal model and the role of sigmoid-like functions in psychology and behavioural science.